\documentclass[preprintnumbers,superscriptaddress,nofootinbib,aps,prd,floatfix]{revtex4}
\usepackage{amsmath,amssymb}
\usepackage{graphicx} 
\usepackage{epstopdf} 
\usepackage{slashed}
\usepackage[caption=false]{subfig}
\usepackage{xcolor}
\usepackage{multirow}
\usepackage{hyperref}
\usepackage{snapshot}
\usepackage[scientific-notation=true,round-mode=places,round-precision=1]{siunitx}

\usepackage{footmisc}
\renewcommand{\thefootnote}{\fnsymbol{footnote}}

\usepackage{feynmp-auto}

\renewcommand{\arraystretch}{1}
\numberwithin{equation}{section}

\definecolor{rossos}{rgb}{0.8,0.2,0.3}
\definecolor{bluscuro}{rgb}{0.15, 0.2, .85}
\definecolor{bluchiaro}{cmyk}{1,.3,0.,0.1}

\definecolor{orange}{rgb}{1,0.5,0}
\definecolor{blue}{rgb}{0,0,1}

\hypersetup{colorlinks, citecolor=bluscuro, linkcolor=bluscuro, urlcolor=bluscuro}

\makeatother   

\def\tm{\widetilde m}

 \def\be   {\begin{equation}}   \def\ee   {\end{equation}}
 \def\ba   {\begin{array}}      \def\ea   {\end{array}}
 \def\bea  {\begin{eqnarray}}   \def\eea  {\end{eqnarray}}
 \def\bean {\begin{eqnarray*}}  \def\eean {\end{eqnarray*}}

\begin{document}

\title{$VHH$ production at the High-Luminosity LHC}

\author{Karl Nordstr\"om} \email{k.nordstrom@nikhef.nl}
\affiliation{Nikhef, Science Park 105, NL-1098 XG Amsterdam, The Netherlands\\[0.1cm]}
\affiliation{Laboratoire de Physique Theorique et Hautes Energies (LPTHE),
UMR 7589 CNRS \& Sorbonne Universit\'e, 4 Place Jussieu, F-75252, Paris, France\\[0.1cm]}
\author{Andreas Papaefstathiou} \email{apapaefs@cern.ch}
\affiliation{Nikhef, Science Park 105, NL-1098 XG Amsterdam, The Netherlands\\[0.1cm]}
\affiliation{Institute for Theoretical Physics Amsterdam and Delta
  Institute for Theoretical Physics, University of Amsterdam, Science
  Park 904, 1098 XH Amsterdam, The Netherlands.\\[0.1cm]}
\date{\today}                                           

\preprint{Nikhef 2018-028}

\begin{abstract}
We study the phenomenology of associated production of a vector boson with a pair of Higgs bosons ($VHH$) at the High-Luminosity LHC (HL-LHC). Despite the low rate of this channel, the scaling of the cross section suggests a measurement could be a useful probe of modifications of the trilinear Higgs boson coupling and anomalous interactions in the gauge-Higgs sector. We focus on both $WHH$ and $ZHH$ production, using the leptonic ($W \to l \nu$, $Z \to ll$, $Z \to \nu \nu$) decay modes of the vector bosons and the $HH \to 4b$ di-Higgs decay mode. We show that top pair backgrounds are problematic for the $W \to l \nu$ and $Z \to \nu \nu$ channels, leaving $Z \to ll$ as the most promising decay mode. However, even for this channel, we find limited sensitivity due to a low signal rate. We discuss some potential avenues for improvement.
\end{abstract}

\maketitle

\section{Introduction}

Since the discovery of a scalar resonance with the properties of a Standard Model-like Higgs boson by the ATLAS and CMS collaborations in 2012 \cite{Aad:2012tfa,Chatrchyan:2012xdj}, the focus of the experiments and phenomenological community has shifted towards precise measurements of the properties of this new particle and its interactions \cite{Khachatryan:2016vau}. Of particular interest is the trilinear self-interaction coupling, $\lambda$, since this would provide a first model-independent measurement of the shape of the scalar potential, that could be related, e.g., to models of strong first-order phase transition necessary for baryogenesis \cite{Reichert:2017puo}.

At colliders the focus has been on direct di-Higgs boson production as the measurement channel of choice. We will continue this tradition here. We note, however, that single Higgs production channels are also sensitive to the trilinear coupling at one-loop \cite{McCullough:2013rea,Degrassi:2016wml,DiVita:2017eyz} and these may turn out to be competitive with direct production, especially at future lepton colliders where single Higgs-strahlung can provide the most sensitive channel overall.

The leading $gg \to HH + X$ channel of di-Higgs production at the LHC has been studied in a range of final states in the phenomenological literature \cite{Baur:2003gpa,Baur:2003gp,Dolan:2012rv,Papaefstathiou:2012qe,Baglio:2012np,Barger:2013jfa,Barr:2013tda,Maierhofer:2013sha,Goertz:2013kp, Frederix:2014hta,Goertz:2014qta,Barr:2014sga,deLima:2014dta,Wardrope:2014kya,Azatov:2015oxa,Behr:2015oqq, Papaefstathiou:2015iba, Papaefstathiou:2017hsb}, and by ATLAS and CMS \cite{ATL-PHYS-PUB-2016-024,ATL-PHYS-PUB-2017-001,Aaboud:2018knk,CMS-PAS-HIG-16-026,Sirunyan:2017djm,Sirunyan:2017guj}. Additionally di-Higgs final states have been studied in associated production with a top quark pair \cite{Englert:2014uqa,Liu:2014rva,ATL-PHYS-PUB-2016-023} and in associated production with two jets (which includes the leading vector boson fusion contribution) \cite{Dolan:2013rja,Dolan:2015zja,Nakamura:2016agl,Bishara:2016kjn}. In this paper we focus on associated production with a weak boson which has previously been studied at hadron colliders \cite{Barger:1988jk,Moretti:2010kc,Cao:2015oxx,Li:2016nrr,Li:2017lbf} and lepton colliders \cite{Maltoni:2018ttu}. Our objective is to perform a realistic sensitivity analysis of this production channel to modifications of the trilinear Higgs coupling $\lambda$ and anomalous quartic gauge-Higgs interactions of the form $V V HH$. Foreshadowing the results of our analysis, we choose to work in a simple ``anomalous'' coupling framework, where the Standard Model (SM) couplings are modified by simple rescaling: $\lambda = \lambda_{SM}(1 + c_3)$, where $\lambda_{SM}$ is the SM value of the triple Higgs boson coupling and $c_3$ parametrises the modifications coming from new physics. Equivalently, we separately consider modifications of the quartic gauge-Higgs couplings through $g_{VVHH} = g_{VVHH,\mathrm{SM}} (1 + c_{VVHH})$ for $V=\{Z,W\}$.

\begin{figure}[h]
\hspace*{-.8cm}\begin{tabular}{ccc}
\begin{fmffile}{diag1}
\begin{fmfgraph*}(100,100)
\fmfleft{i1,i2}
\fmfright{o1,o2,o3}
\fmf{fermion,tension=2}{i1,v1}
\fmf{fermion,tension=2}{v1,i2}
\fmf{photon,tension=2}{v1,v2}
\fmf{dashes}{v2,v3}
\fmf{dashes}{v3,o1}
\fmf{dashes}{v3,o2}
\fmf{photon}{v2,o3}
\end{fmfgraph*}
\end{fmffile}
&
\begin{fmffile}{diag2}
\begin{fmfgraph*}(100,100)
\fmfleft{i1,i2}
\fmfright{o1,o2,o3}
\fmf{fermion,tension=2}{i1,v1}
\fmf{fermion,tension=2}{v1,i2}
\fmf{photon,tension=2}{v1,v2}
\fmf{dashes}{v2,o1}
\fmf{dashes}{v2,o2}
\fmf{photon}{v2,o3}
\end{fmfgraph*}
\end{fmffile}
&
\begin{fmffile}{diag3}
\begin{fmfgraph*}(100,100)
\fmfleft{i1,i2}
\fmfright{o1,o2,o3}
\fmf{fermion,tension=2}{i1,v1}
\fmf{fermion,tension=2}{v1,i2}
\fmf{photon,tension=2}{v1,v2}
\fmf{dashes}{v2,o1}
\fmf{photon}{v2,v3}
\fmf{dashes}{v3,o2}
\fmf{photon}{v3,o3}
\end{fmfgraph*}
\end{fmffile}
\end{tabular}
\caption{The diagrams that contribute to $pp \to VHH$ at tree level. The third one is accompanied by a '$u$ channel' one with $p_{H_1} \leftrightarrow p_{H_2}$ which we do not draw.}\label{fig:diagrams}
\end{figure}
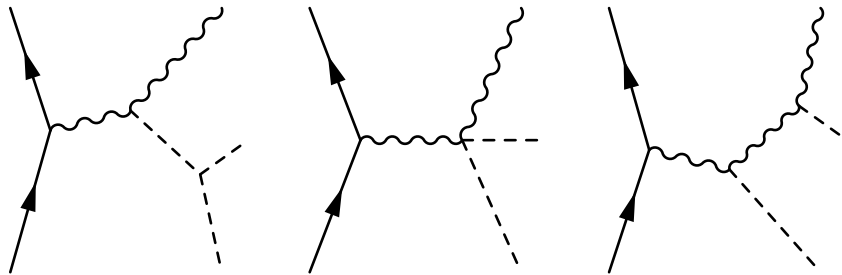

\begin{figure}[h]
\centering
 \includegraphics[width=.49\textwidth]{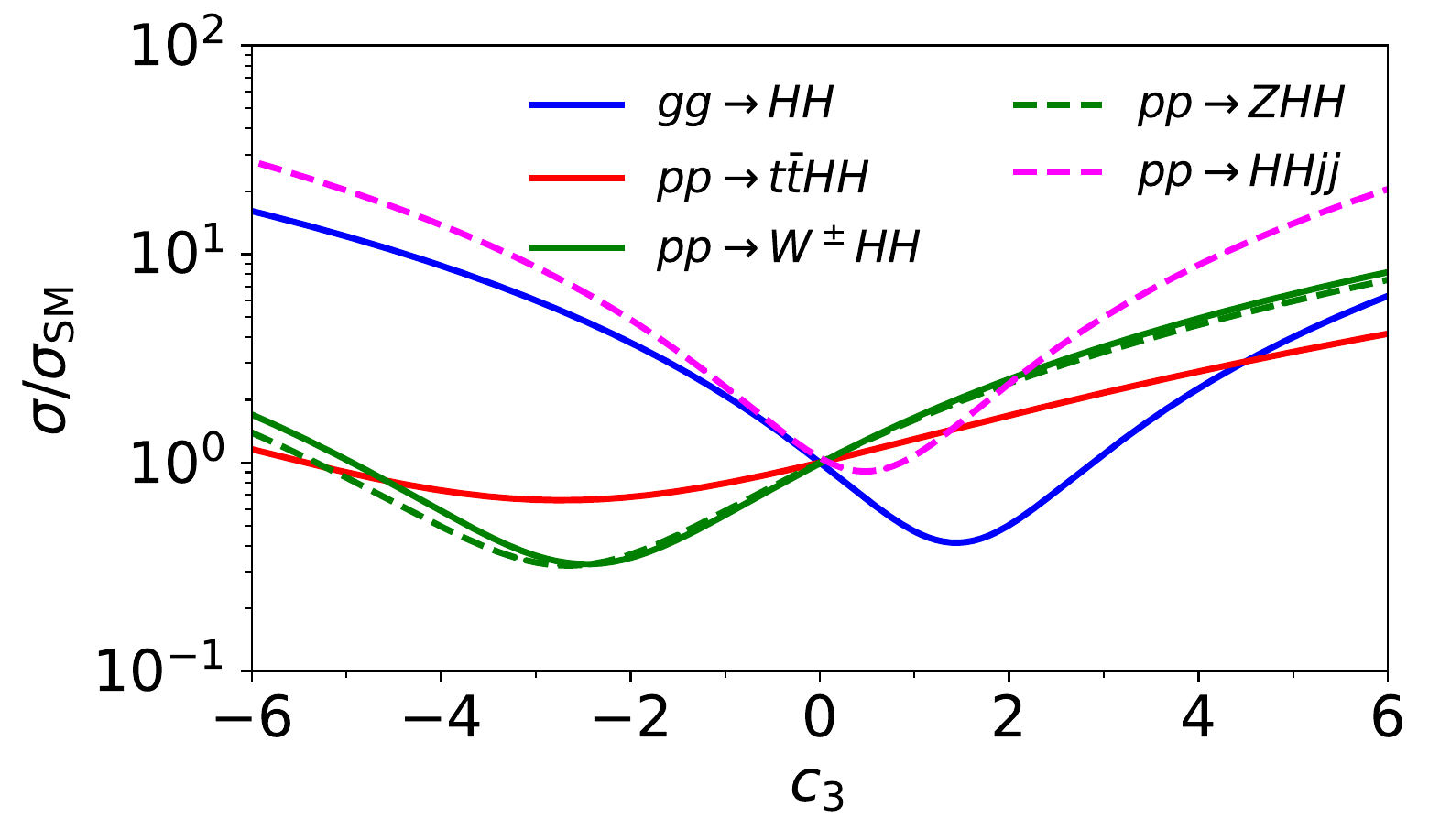}
\caption{The scaling of the cross section with modifications to $\lambda = \lambda_{SM}(1 + c_3)$ of the leading di-Higgs production channels. The $VHH$ channels are at this level the most sensitive to small positive modifications to $\lambda$. \label{fig:c3_scaling}}
\end{figure}

The diagrams which contribute to this process at tree level are presented in Figure~\ref{fig:diagrams}. As noted in \cite{Cao:2015oxx}, the interference between the $\lambda$ contribution and the other diagrams in this channel is such that it could potentially offer a sensitive probe of $\lambda > \lambda_\textrm{SM}$. This can be most easily seen if we amputate the quarks to focus on the $V^\mu \to V^\nu H H$ subdiagrams setting the gauge boson mass to equal the Higgs boson mass, $m_V = m_H$, for simplicity, where we find at threshold:

\begin{equation}
\mathcal{M}^{\mu \nu} = \frac{2 g^{\mu \nu} m_H^2}{3 v^2} [7 + 3 (1+c_3) ]\;,
\end{equation}
where $\lambda = (1+c_3) \lambda_{SM}$. This suggests the interference pattern is such that the cross section is smallest for $\lambda \approx -2 \lambda_\textrm{SM}$, in contrast to $gg \to HH$ and vector boson fusion $qq \to qq HH$ where this occurs for $\lambda > \lambda_{SM}$, as demonstrated in Figure~\ref{fig:c3_scaling}.

Another potentially interesting feature of $VHH$ production is that it gives us sensitivity to the $WWHH$ and $ZZHH$ vertices, which could be modified in strongly coupled and extra dimensional models of electroweak symmetry breaking \cite{Csaki:2015hcd}. These are probed very efficiently by vector boson fusion $qq \to qq HH$ production \cite{Dolan:2013rja,Dolan:2015zja,Bishara:2016kjn} since the contributing diagrams are such that the high di-Higgs invariant mass ($m_{HH}$) region is only suppressed by valence quark parton density functions which leads to large sensitivity when using boosted reconstruction techniques. However this channel is only sensitive to a linear combination of the $WWHH$ and $ZZHH$ vertices, whereas the ability to tag the vector boson in $VHH$ production could potentially allow us to constrain these independently as $ZHH$ and $WHH$ production are sensitive to $g_{ZZHH}$ and $g_{WWHH}$ separately at leading order. A similar observation motivated the study of $WWH$ production in \cite{Englert:2017gdy}.

We will work at $\sqrt{s} = 13$~TeV proton-proton collision energy throughout the paper. Since the signal processes are $q \bar{q}$-initiated at tree-level the cross sections grow more slowly when going to higher collider energies as compared to gluon fusion-initiated di-Higgs production channels; for this reason the qualitative changes that can be expected by considering, e.g., a 100 TeV proton-proton collider will be limited to a potentially larger data samples and detector improvements.

\section{Setup of calculation and analysis}

\label{sec:setup}

We use the \textsc{MadGraph5\_aMC@NLO} framework \cite{mg5} to generate both signal and background events at leading or next-to-leading QCD order \cite{Ossola:2007ax,Binoth:2008gx,Mastrolia:2012bu,Alwall:2014hca,Peraro:2014cba,Hirschi:2016mdz,Denner:2016kdg} depending on the number of legs of the process. We use \textsc{MadSpin} \cite{Artoisenet:2012st} to decay particles with the correct spin correlations (reweighting the branching ratio to the HXSWG recommendation for $m_H = 125.0$ GeV in \cite{Heinemeyer:2013tqa}) and \textsc{Herwig} 7 \cite{Bahr:2008pv, Gieseke:2011na, Arnold:2012fq, Bellm:2013lba, Bellm:2015jjp, Bellm:2017bvx} to shower and hadronise the parton-level events. We employ the \textsc{Rivet} framework \cite{Buckley:2010ar} to analyse the hadron level events.

We use the default $b$-tagging implementation in \textsc{Rivet} which ``ghost-associates'' \cite{Cacciari:2007fd,Cacciari:2008gn} $b$ and $c$ mesons to the jets to define $b$ and $c$ jets, which are then assigned $b$-tags with an efficiency of 77\% for $b$ jets, falling to (100/6)\% for $c$ jets and (100/134)\% for light jets, corresponding to a standard operating point for the ATLAS MV2c10 algorithm \cite{ATL-PHYS-PUB-2016-012}.

\section{$ZHH$ Production}

$ZHH$ production does not have an a priori obviously superior candidate between the invisible ($Z \to \nu \nu$) and and leptonic ($Z \to l l$) decay channels of the $Z$ due to the lower branching ratio to charged leptons. The cross section before branching ratios at NNLO QCD \footnote{The relatively large NNLO/NLO $K$-factor of 1.2 is caused by the introduction of $gg \to ZHH$ contributions at NNLO. Due to the low total cross section of the $pp \rightarrow ZHH$ process, we only approximately consider this channel through the $K$-factor and do not include it explicitly in our Monte Carlo simulations.} is $\sim$0.37 fb \cite{Djouadi:2012zc,Li:2017lbf}. Previous phenomenological studies of $ZHH$ production at parton level have employed the invisible decay channel \cite{Cao:2015oxx},\footnote{As far as we understand, no parton showering was employed in the study of~\cite{Cao:2015oxx}.} and it also forms the most sensitive $ZH$ channel in the recent $H \to b \bar{b}$ analysis by ATLAS \cite{Aaboud:2017xsd}. These exploit the fairly soft missing energy spectrum of top pair backgrounds, leaving $Z+$heavy flavour backgrounds dominant while gaining statistics from the relatively large $Z \to \nu \nu$ branching ratio. For this study we have investigated both of these $Z$ decay channels.

\subsection{$Z \to \nu \nu$}

The $Z \to \nu \nu$ channel is attractive due to the relatively large branching ratio:

\begin{equation}
\frac{\text{Br}\left( Z \to \nu \nu \right)}{\text{Br}\left( Z \to e^+ e^-, \mu^+ \mu^- \right)} \sim \frac{20\%}{6.7\%} \sim 3
\end{equation}

\begin{figure}[t]
\centering
\subfloat[]{
 \includegraphics[width=.49\textwidth]{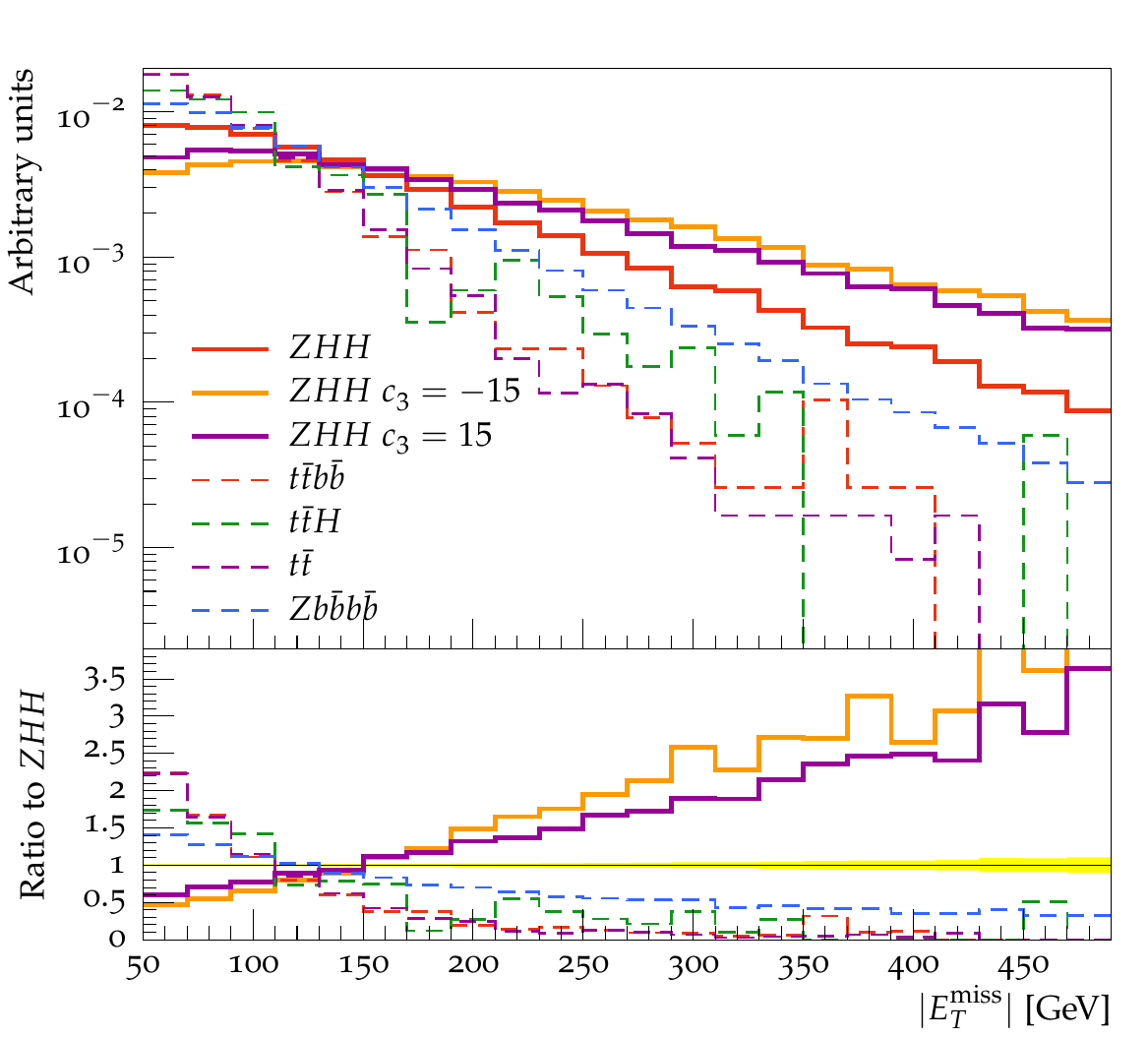}
 \label{fig:zhh_met}
}
\subfloat[]{
 \includegraphics[width=.49\textwidth]{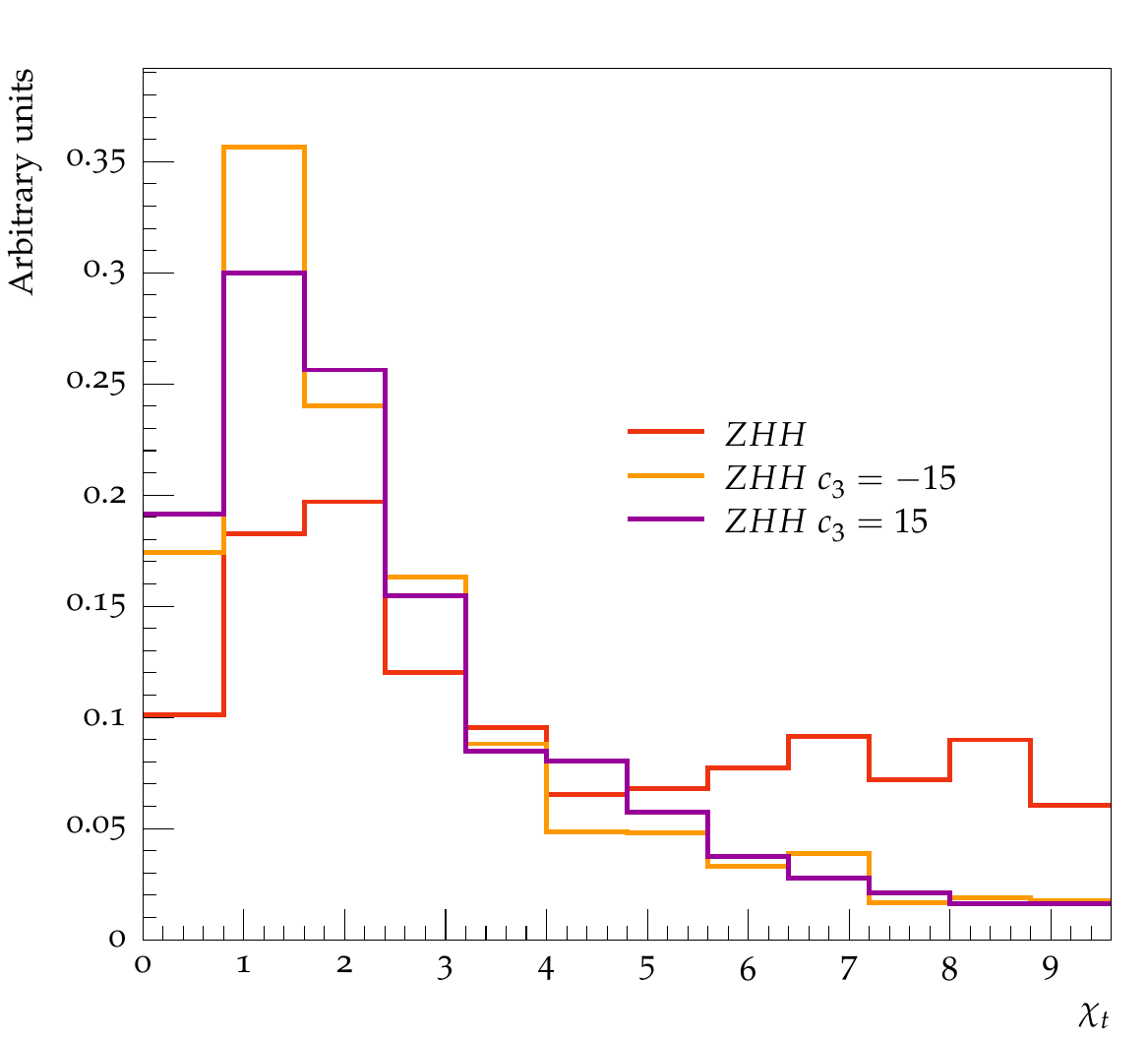}
 \label{fig:zhh_top_fit}
}
\caption{Shape comparison of the missing transverse energy distributions of the signal and leading backgrounds (left) and the $\chi_t$ distribution after all other selections for the signal for three different values of $\lambda$ (right) in the $Z \to \nu \nu$ analysis. $c_3$ is defined as $\lambda = \lambda_{SM}(1+c_3)$. \label{fig:zhh_nunu_kinematics}}
\end{figure}

This gives a $ZHH$ $(Z \to \nu \nu, HH \to 4b)$ NNLO QCD cross section of $\sim$0.025 fb. Additionally, aggressive cuts on the missing transverse energy $|E_T^{\textrm{miss}}|$ and vetoes on identified leptons allows multijet and top pair backgrounds to be controlled enough for this channel to be the most sensitive in the $ZH$ ($H \to b \bar{b}$) context \cite{Aaboud:2017xsd}. We do not take into account the $|E_T^{\textrm{miss}}|$ trigger efficiency in our analysis: according to current ATLAS performance~\cite{Aaboud:2016leb}, this would lower our cross sections by a factor of $\sim 2$ due to the cut of 100 GeV. However, one can expect this to change in the future HL-LHC runs. Unfortunately we find that top quark pair (associated) production is a very challenging background for the $Z \to \nu \nu$ channel of $ZHH$ ($HH \to 4b$) production. Following the strategy of the $ZH$ analysis, we consider further avenues to control it:

\begin{itemize}
\item Vetoing on prompt leptons requires them to be hard enough to be identified and within the detector volume. Defining veto-able leptons as those with $p_T > 5$ GeV within the inner detector $|\eta| < 2.5$ (similar to the lepton veto used in monojet analyses, see, e.g.~\cite{Aad:2015zva}) with perfect identification efficiency (even for $\tau$ leptons) improves the signal-to-background ratio considerably. However some 9\% of semi-leptonically decaying top pairs in $t \bar{t}$ production (the fraction varies only slightly for $t \bar{t} b \bar{b}$ and $t \bar{t} H$ production) still pass and in fact form the dominant background.
\item The missing energy spectrum is softer for the top backgrounds than the signal. However due to the already low signal cross section we keep the missing transverse energy cut at a relatively low value, $|E_T^{\textrm{miss}}| > 100$ GeV. Further tuning of this value could potentially improve the sensitivity of the analysis. The $|E_T^{\textrm{miss}}|$ distribution for the signal at three different values of $\lambda$ and the leading backgrounds are presented in Figure~\ref{fig:zhh_met}.
\item We require the kinematics of the four $b$-tagged jets $i,j,i',j'$ in the event to resemble those of a pair of Higgs decays by minimising $\chi_{HH} =\sqrt{ \left( \frac{m_{ij} - m_H}{ 0.1 m_H} \right)^2 + \left( \frac{m_{i'j'} - m_H}{ 0.1 m_H} \right)^2 }$.
\item We further attempt to reduce top backgrounds by finding the three jet permutation $t^*$ which most closely reconstructs the three jet mass $m_t = 172.5$ GeV with a sub-permutation $W^*$ which reconstructs the two jet mass $m_W = 80.4$ GeV at the end of the analysis by minimising $\chi_t = \sqrt{ \left( \frac{m_{t^*} - m_t}{ 0.1 m_t} \right)^2 + \left( \frac{m_{W^*} - m_W}{ 0.1 m_W} \right)^2 }$. However we ultimately find that it is difficult to use this to improve the sensitivity since the scales in the signal (with 4 jets from two Higgs decays) make it easy to fake a top candidate, in particular when the cross section is dominated by diagrams containing the trilinear coupling, as shown in Figure~\ref{fig:zhh_top_fit}.
\end{itemize}

The $Z \to \nu \nu$ analysis steps are as follows:
\begin{enumerate}
\item Require \textit{no} identified leptons with $p_T > 5$ GeV inside $|\eta| < 2.5$.
\item Require $|E_T^{\textrm{miss}}| > 100$ GeV.
\item Require at least 4 jets with $p_T > 40$ GeV.
\item Require the 4 leading jets to be $b$-tagged using the definition and efficiencies detailed in Section~\ref{sec:setup}.
\item Require that these $b$-tagged jets have the kinematics of a $HH$ pair decay, $\chi_{HH} < 1.6$.
\end{enumerate}

\subsection{$Z \to ll$}
\label{sec:zll}

\begin{figure}[t]
\centering
\subfloat[]{
 \includegraphics[width=.49\textwidth]{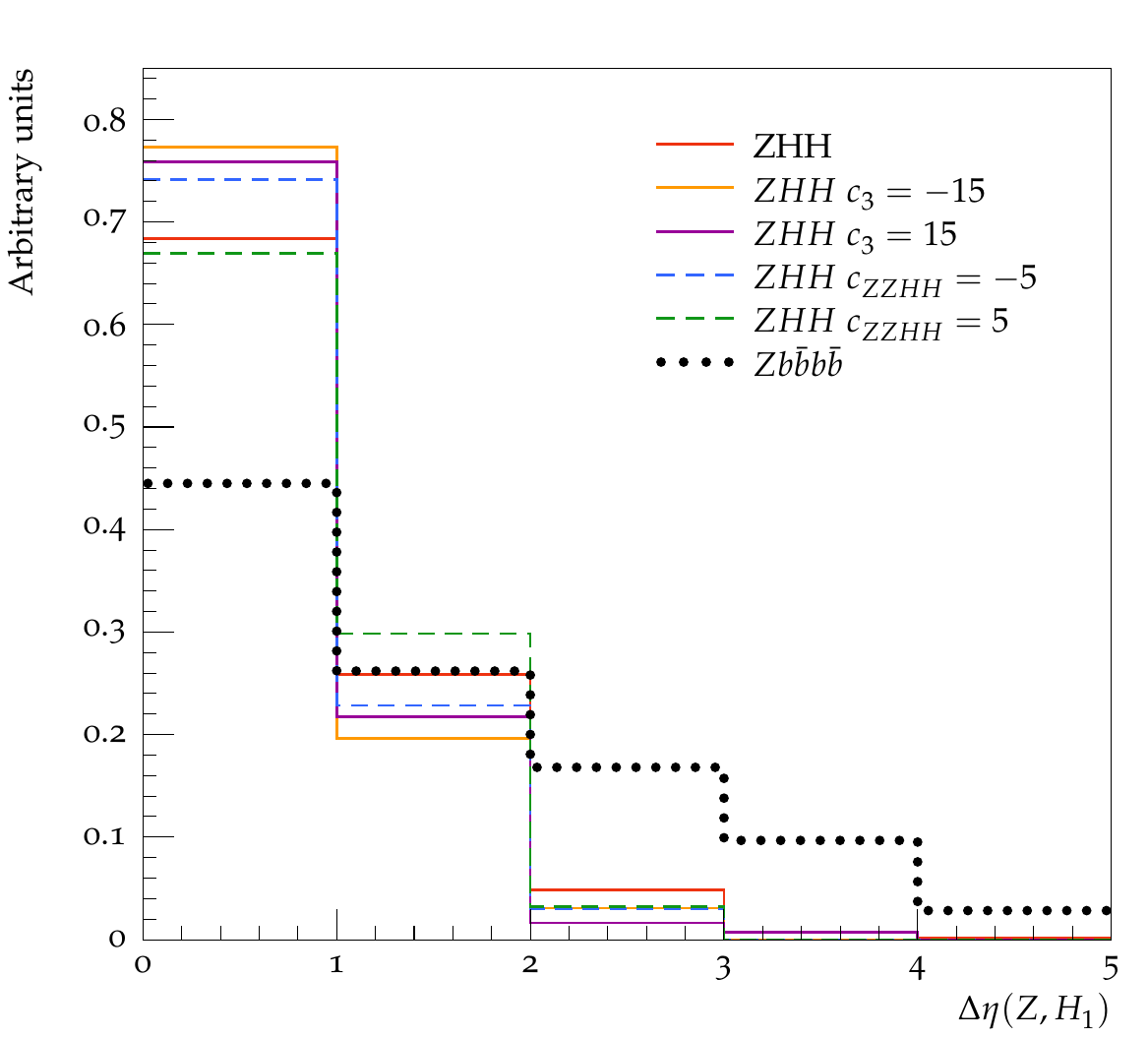}
 \label{fig:zhh_deta_v_h1}
 }
\subfloat[]{
 \includegraphics[width=.49\textwidth]{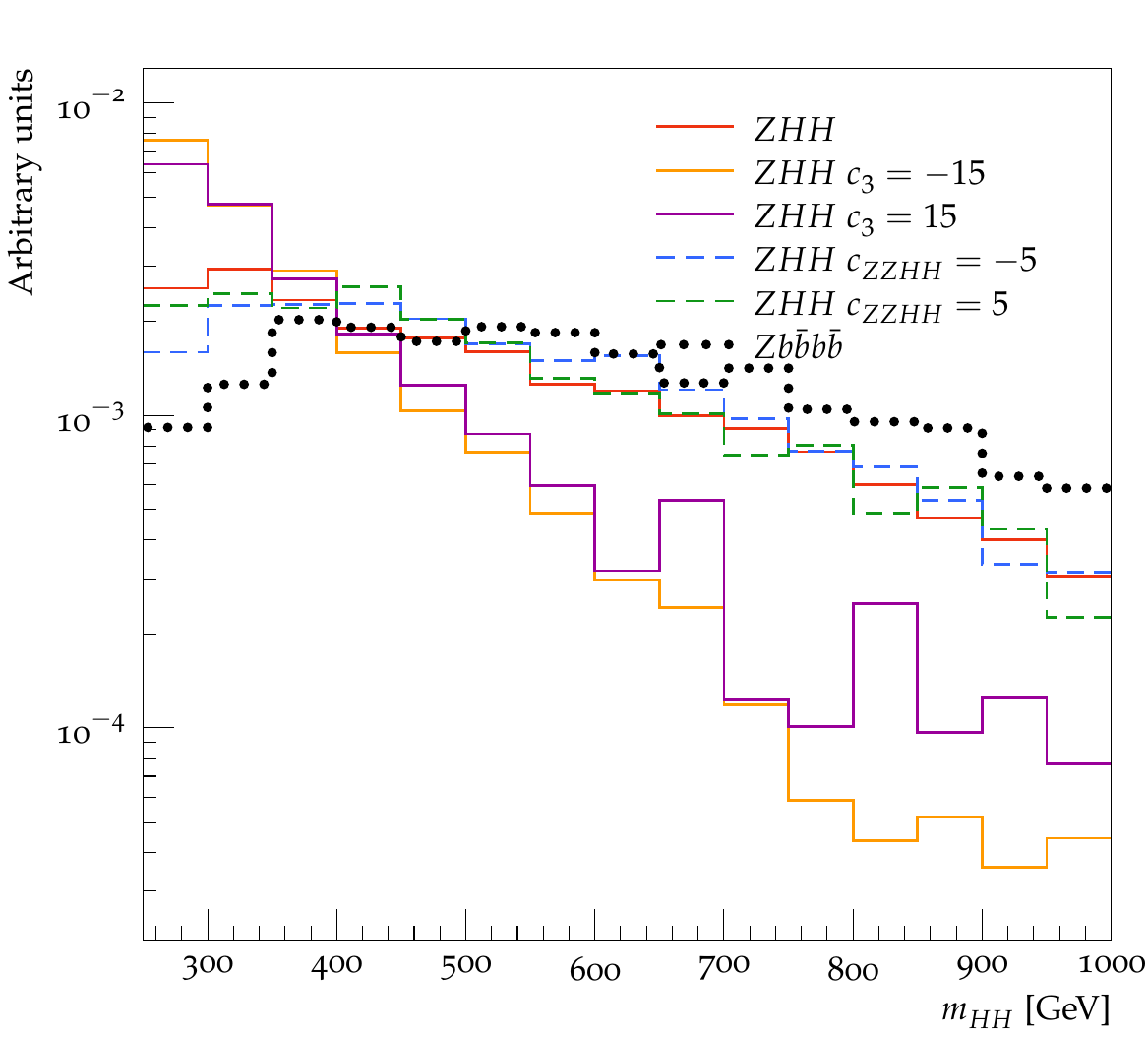}
 \label{fig:zhh_mhh}
 }
\caption{Shape comparison of $\Delta \eta(Z,H_1)$ (left, where $H_1$ is the leading Higgs candidate in $p_T$) and $m_{HH}$ (right) for the signal and the dominant background $Zb\bar{b}b\bar{b}$ after all other selections in the $Z \to ll$ analysis at five different values of $c_3$ and $c_{ZZHH}$.}
\end{figure}

The $Z \to ll$ ($l = e, \mu$) channel, when compared to $Z \to \nu \nu$ above, suffers from a smaller branching ratio giving an NNLO QCD parton level cross section for $ZHH$ $(Z \to ll, HH \to 4b)$ of $\sim$0.008~fb. However it allows for top backgrounds to be controlled through a combination of aggressive cuts on $m_{ll}$ which can be justified due to the excellent lepton momentum resolution of the LHC experiments, and requiring $|E_T^{\textrm{miss}}| < 50$ GeV. Since we find that $Z$ + heavy flavour backgrounds are dominant and are able to reconstruct the $Z$ boson completely we have investigated angular observables in the $Z$, $H_1$, and $H_2$ systems (where $H_{1,2}$ are the leading and sub-leading reconstructed Higgs candidate in $p_T$, respectively) and find that $\Delta \eta(Z,H_1)$ can be used to significantly reduce this background at little signal cost, see Figure~\ref{fig:zhh_deta_v_h1}. Other angular observables may also carry some additional information, however due to the low signal rates a multivariate approach would be required to make use of this (while also balancing the non-negligible top pair backgrounds).

The $Z \to ll$ analysis steps are as follows:
\begin{enumerate}
\item Require exactly two same flavour opposite charge electrons or muons inside $|\eta| < 2.5$ with $p_T > 25$ GeV.
\item Require these leptons to have an invariant mass compatible with that originating from a $Z$ boson decay, $|m_{ll} - m_Z| < 5$ GeV.
\item Require $|E_T^{\textrm{miss}}| < 50$ GeV.
\item Require at least 4 jets with $p_T >$ 40 GeV. Veto event if any of these overlap with a lepton.
\item Require the 4 leading jets to be $b$-tagged using the definition and efficiencies detailed in Section~\ref{sec:setup}.
\item Require that these $b$-tagged jets have the kinematics of a Higgs boson pair pair decay, $\chi_{HH} < 1.6$.
\item Require that the leading Higgs candidate $H_1$ and the reconstructed $Z$ boson are not too far separated in pseudorapidity, $\Delta \eta(Z,H_1) < 2$.
\end{enumerate}

The visible cross sections after these selections are applied for the signal and backgrounds are presented in Table~\ref{tab:xstable_zhh}. Using our selections the two channels end up being competitive with each other: the $Z \to \nu \nu$ channel has higher signal statistics but a low signal-to-background ratio $S/B \sim 1/660$, while the $Z \to ll$ channel has lower statistics but a higher $S/B \sim 1/85$.

Techniques making use of boosted topologies have been shown to offer sensitivity improvements in the $HH \to 4b$ final state in $HH$ \cite{deLima:2014dta} and $HHjj$ \cite{Bishara:2016kjn} production, by making use of lower background yields in the high-$m_{HH}$ tail of the distribution and the ability to separate signal from background using boosted reconstruction techniques. In particular the sensitivity of $HHjj$ measurements to modifications of $c_{VVHH}$ is greatly enhanced in the VBF topology.  We have checked that the $m_{HH}$ distribution after all other cuts in the $Z \to ll$ analysis\footnote{The kinematics of the Higgs boson system in all three channels are similar, so this conclusion also applies to the $Z \to \nu \nu$ and $W \to l \nu$ analyses.} itself only contains new information in the low $m_{HH}$ region compared to the dominant $Zb \bar{b} b \bar{b}$ background when taking the kinematic dependence of $c_{3}$ and $c_{VVHH}$ into account, see Figure~\ref{fig:zhh_mhh}. We note that the behaviour when the process is dominated by the two different types of vertices we consider can be explained with reference to the contributions detailed in Figure~\ref{fig:diagrams}: at tree level the $c_{VVHH}$ and $c_{VVH}^2$ contributions scale as $\sim v^2/m_{HH}^2$ at large $m_{HH}$, while the $\lambda$ contribution scales as $\sim v^4/m_{HH}^4$. This explains the similarity in scaling in the differential cross section between the Standard Model and when it is dominated by the $c_{VVHH}$ term, and the faster falloff when it is dominated by the $\lambda$ term. This would allow a multivariate analysis to improve our results for large modifications to $\lambda$, but is difficult to make use of in a cut-based analysis.

\section{$WHH$ Production}

The production of a pair of Higgs bosons in association with a $W$ boson, $WHH$, has the advantage of a larger cross section compared to $ZHH$: $\sigma (WHH \rightarrow \l \nu_l + 4b) \simeq 0.04$~fb at NNLO QCD ($l = e, \mu$) \cite{Li:2016nrr}. The main challenge is to distinguish this channel from $t\bar{t}+X$ backgrounds, where $X$ can be jets (including $b\bar{b}$), or a Higgs boson decaying to $b\bar{b}$.

We follow a similar analysis strategy to~\cite{Cao:2015oxx}, which consists of the following steps:

\begin{enumerate}
\item Require exactly one lepton inside $|\eta| < 2.5$ with $p_T > 25$ GeV.
\item Require $|E_T^{\textrm{miss}}| > 40$ GeV.
\item Require $m_T \leq m_W$ and $H_T \geq 400$~GeV, where $m_T=\sqrt{ 2 p_T^\ell |E_T^{\textrm{miss}}| (1 - \cos \phi)}$, $\phi$ the azimuthal angle between the lepton and the missing energy vector, and $H_T$ is the scalar sum of the transverse momenta of jets and the charged lepton.
\item Require at least four jets and that the lepton does not lie within $\Delta R < 0.4$ of any jet.
\item Require the 4 leading jets to be $b$-tagged using the definition and efficiencies detailed in Section~\ref{sec:setup}.
\item Require that these $b$-tagged jets have the kinematics of a $HH$ pair decay, $\chi_{HH} < 1.6$.
\end{enumerate}

Similar to the $Z \to \nu \nu$ analysis we also attempt to reduce the top quark backgrounds using the observable $\chi_t$ defined above. However, since the $VHH$ kinematics are very similar between $VHH$ channels, we find the same situation as in Figure~\ref{fig:zhh_top_fit} and, consequently, we do not employ $\chi_t$ here either.

\begin{turnpage}
\begin{table*}[!h]
  \hspace{-0.5cm}  \begin{tabular}{lccccccccc}
    \toprule
    Cut ($Z \to ll$)     		         &  $ZHH$  &  $Zb\bar{b}b\bar{b}$ &  $Zb\bar{b}c\bar{c}$ &  $Zt\bar{t}$ &  $ZZb\bar{b}$    & $t\bar{t}H$   & $t\bar{t}b\bar{b}$  & $t\bar{t}c\bar{c}$ & $t\bar{t}$ \\
    \botrule
    2 same flavour leptons	         &  \num{5.19E-06}&   \num{4.02E-03}&  \num{7.48E-02}  & 	\num{2.78E-02} & \num{3.05E-03} & \num{3.82E-03} & \num{5.06E-02}  &  \num{5.89E-02} &  \num{1.15E+01}	\\
    $|m_{ll} - m_Z| < 5$ GeV	       &  \num{4.35E-06}&   \num{3.39E-03}&  \num{6.33E-02}  &  \num{2.02E-02} & \num{2.62E-03} &	\num{2.55E-04} & \num{3.60E-03} & 	\num{4.02E-03} &  \num{8.71E-01}	\\
    $|E_T^{\textrm{miss}}| < 50$ GeV &  \num{4.35E-06}&   \num{3.39E-03}&   \num{6.33E-02} & 	\num{1.50E-02} & \num{2.62E-03} &	\num{7.15E-05} & \num{1.06E-03} & 	\num{1.03E-03} &  \num{3.63E-01}	\\
    $\ge 4$ jets with $p_T > 40$ GeV &  \num{1.28E-06}&   \num{5.93E-04}&   \num{2.12E-03} & 	\num{7.85E-03} & \num{7.43E-05} &	\num{2.47E-05} & \num{2.23E-04} & 	\num{2.78E-04} &  \num{3.43E-02}	\\
    4 leading jets $b$-tagged	       &  \num{1.42E-07}&   \num{5.13E-05}&   \num{4.99E-06} & 	\num{5.71E-06} & \num{3.89E-06} & \num{2.17E-06} & \num{1.53E-05} & 	\num{2.00E-06} &  \num{2.88E-06}	\\
    $\chi_{HH} < 1.6$	               &  \num{6.84E-08}&   \num{3.45E-06}&   \num{3.58E-07} & 	\num{9.94E-07} & \num{3.84E-07} &	\num{4.24E-08} & \num{1.20E-06} & 	\num{1.82E-07} &  \num{2.09E-07}	\\
    $\Delta \eta (Z, H_1) < 2$	     &  \num{6.44E-08}&   \num{2.43E-06}&   \num{2.97E-07} & 	\num{8.11E-07} & \num{3.44E-07} &	\num{4.23E-08} & \num{1.20E-06} & 	\num{1.82E-07} &  \num{2.09E-07}	\\
    \botrule
    Events in 3 ab$^{-1}$            & \num{0.1932}     & \num{7.283}   & \num{0.8921}     & \num{2.434}     & \num{1.033} &  \num{0.1269} &     \num{3.602}      & \num{0.5463}  &  \num{0.6268}
    \\
    \\
    \\
    \toprule
    Cut ($Z \to \nu \nu$)     		   &  $ZHH$   &  $Zb\bar{b}b\bar{b}$ &  $Zb\bar{b}c\bar{c}$ &  $Zt\bar{t}$&  $ZZb\bar{b}$ & $t\bar{t}H$  & $t\bar{t}b\bar{b}$   & $t\bar{t}c\bar{c}$ & $t\bar{t}$  \\
    \botrule
    No identified leptons         	 &  \num{2.19E-05}& \num{1.77E-02} &  \num{3.89E-01}    &\num{4.76E-02} &\num{1.45E-02} &\num{1.09E-01}& 	\num{1.05E+01} & 	\num{1.73E+00} & 	\num{3.61E+02} 	\\
    $|E_T^{\textrm{miss}}| > 100$ GeV&  \num{1.09E-05}& \num{6.32E-03}&   \num{6.73E-02}    &\num{7.78E-04} &\num{2.70E-03} &\num{1.60E-03}& 	\num{1.36E-01} & 	\num{2.41E-02} & 	\num{4.67E+00} 	\\
    $\ge 4$ jets with $p_T > 40$ GeV &  \num{3.96E-06}& \num{1.55E-03} &  \num{5.66E-03}    &\num{4.74E-04} &\num{2.08E-04} &\num{1.17E-03}& 	\num{4.51E-02} & 	\num{1.29E-02} & 	\num{1.14E+00} 	\\
    4 leading jets $b$-tagged	       &  \num{4.35E-07}& \num{1.28E-04} &  \num{1.12E-05}    &\num{1.57E-05} &\num{1.24E-05} &\num{2.76E-05}& 	\num{5.78E-04} & 	\num{2.36E-05} & 	\num{9.75E-04} 	\\
    $\chi_{HH} < 1.6$	               &  \num{2.28E-07}& \num{6.76E-06} &  \num{5.36E-07}    &\num{9.77E-07} &\num{8.28E-07} &\num{1.04E-06}& 	\num{1.08E-04} & 	\num{1.53E-06} & 	\num{3.16E-05} 	\\
    \botrule
    Events in 3 ab$^{-1}$            & \num{0.6829}   & \num{20.29}     & \num{1.608}       & \num{2.930}   & \num{2.484}   & \num{3.121}  &  \num{323.9}    &  \num{4.579}    &  \num{94.75}
    \\
    \\
    \\
    \toprule
  Cut ($W \to l \nu$)                          &  $WHH$  &  $t\bar{t}$ & $t\bar{t}b\bar{b}$ & $t\bar{t} H$ & $t\bar{t}Z$ &  $W b\bar{b}b\bar{b} $ & $W t\bar{t}$  \\
  \botrule
$\ge 4$ jets with $p_T > 40$ GeV & \num{1.21655176e-05} &\num{64.2435598954} &\num{0.76195423746} &\num{0.102659949898} &\num{0.0162559386795} &\num{0.000763340903633} &\num{0.0357220855331}  \\
Exactly one lepton & \num{8.7371496e-06} &\num{42.9766323286} &\num{0.500789458762} &\num{0.0677452621276} &\num{0.0103128426593} &\num{0.000528411672264}&\num{0.0235628803317}  \\
$|E_T^{\textrm{miss}}| < 40$~GeV& \num{5.4464352e-06} &\num{25.7953793093} &\num{0.315203321722} &\num{0.043829350212} &\num{0.00670747183691} &\num{0.00029563162069} &\num{0.0133018045803}  \\
$m_T < 80.42$~GeV and $H_T > 400$~GeV  & \num{3.86528e-06} &\num{12.7365674377} &\num{0.199443387525} &\num{0.0295903760832} &\num{0.00437529977711} &\num{0.000188936783756} &\num{0.00861082833789}  \\
4 leading jets $b$-tagged & \num{4.2669232e-07} &\num{0.00590637417613} &\num{0.00501908309131} &\num{0.000736589149262} &\num{0.000126095458525} &\num{2.09691174239e-05} &\num{4.98460283373e-06}  \\
$\chi_{HH} < 1.6$ & \num{2.11114224e-07} &\num{0.000529654857454} &\num{0.000211323995019} &\num{7.94878823791e-05} &\num{4.95526466717e-06} &\num{6.11365236396e-07}&\num{7.35570813738e-07}  \\
  \botrule
Events in 3~ab$^{-1}$ & \num{0.633342672} & \num{1588.96457236} & \num{633.971985057} & \num{238.463647137} & \num{14.8657940015} & \num{1.83409570919} & \num{2.20671244121}
     \end{tabular}
     \caption{Cross sections in picobarns for $ZHH$ ($Z \to ll$ and $Z \to \nu \nu$) and $WHH$ ($W \to l \nu$), and backgrounds after the selections described in the text are applied. Generation-level cuts on the invariant masses of lepton pairs, missing transverse energy, and $p_T$ of jets are employed for some of the backgrounds. Top quark branchings depend on the analysis to allow for the possibility of leptons escaping detection, and to ensure there are two leptons for the $Z \to ll$ analysis. The $ZHH$, $WHH$, $Zt\bar{t}$,  $Zt\bar{t}$, $Wt\bar{t}$, $t \bar{t}H$, and $t \bar{t}$ samples are generated at NLO QCD. The $t \bar{t}$ sample is further reweighted to the NNLO+NNLL QCD cross section \cite{Czakon:2011xx}. The other samples are generated at leading order and the $t\bar{t}b\bar{b}$ sample is reweighted to the NLO QCD cross section \cite{Bredenstein:2009aj,Bevilacqua:2009zn}. }
 \label{tab:xstable_zhh}
\end{table*}

\end{turnpage}

\section{Projected limits}

\begin{figure}[h]
\centering
\subfloat[]{
 \includegraphics[width=.49\textwidth]{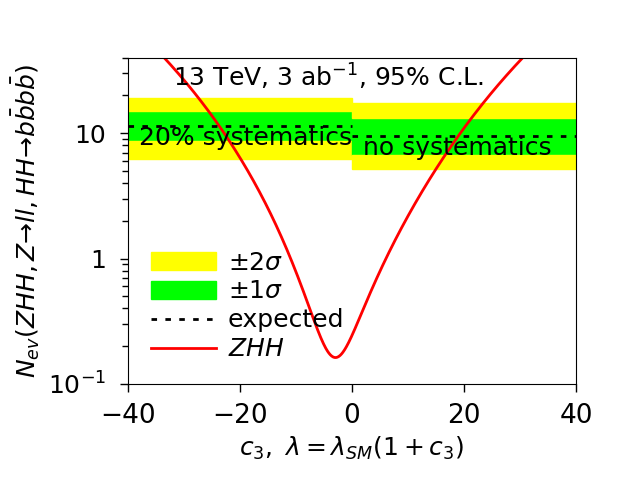}
}
\subfloat[]{
  \includegraphics[width=.49\textwidth]{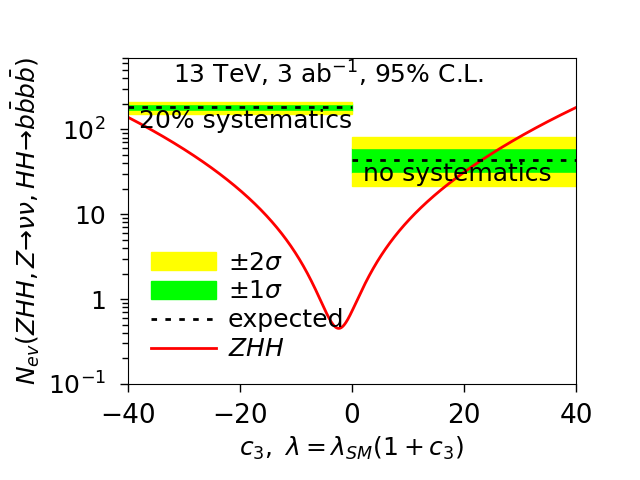}
}
\caption{Projected 95\% confidence level limits on the trilinear Higgs coupling derived from the $ZHH$ analyses ($Z \to ll$ on the left, $Z \to \nu \nu$ on the right) using the full HL-LHC data set. For details on how the signal cross section and visible cross section limits are calculated see text. \label{fig:zhh_limits}}
\end{figure}

\begin{figure}[h]
\centering
\subfloat[]{
 \includegraphics[width=.49\textwidth]{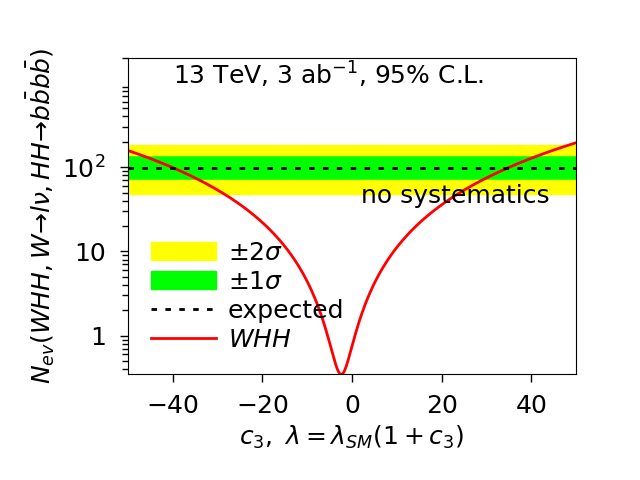}
 \label{fig:whh_limits_c3}
}
\subfloat[]{
  \includegraphics[width=.49\textwidth]{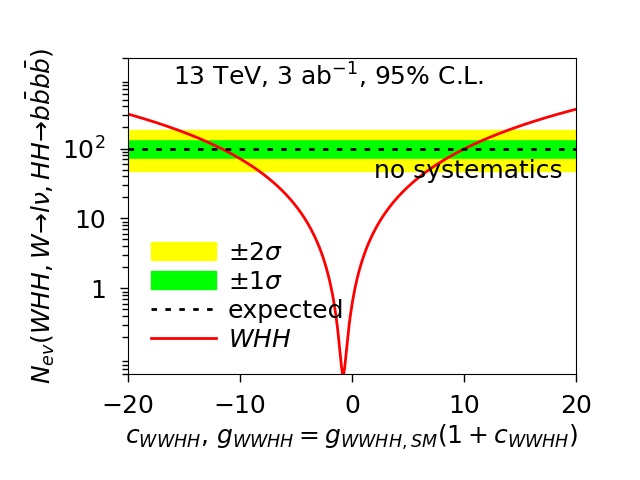}
  \label{fig:whh_limits_cvvhh}
}
\caption{Projected 95\% confidence level limits on the trilinear Higgs coupling (left) and the quartic $WWHH$ coupling (right) derived from the $WHH$ analysis ($W \to l\nu$) using the full HL-LHC data set. Due to the weak sensitivity we do not present results including any systematic uncertainty. For details on how the signal cross section and visible cross section limits are calculated see text. \label{fig:whh_limits}}
\end{figure}

The visible cross sections after each selection in all three analyses are presented in Table~\ref{tab:xstable_zhh}. The projected 95\% confidence level limits on the trilinear Higgs coupling $\lambda$ with the full HL-LHC data set using the three analyses are presented in Figures~\ref{fig:zhh_limits}~and~\ref{fig:whh_limits_c3}: these have been derived by calculating the visible $VHH$ cross section dependence on variations of $\lambda$ (taking the kinematic dependence correctly into account) and fitting this to a polynomial, while the projected cross section constraints are evaluated using Poissonian likelihoods. Results with 20\% systematic uncertainty approximated as a Gaussian uncertainty on the background are also presented to provide an estimate of the sensitivity of these results to realistic experimental conditions. Due to the weak sensitivity of the $WHH$ analysis we do not present results including this systematic uncertainty for this channel.

\begin{figure}[h]
\centering
\subfloat[]{
 \includegraphics[width=.49\textwidth]{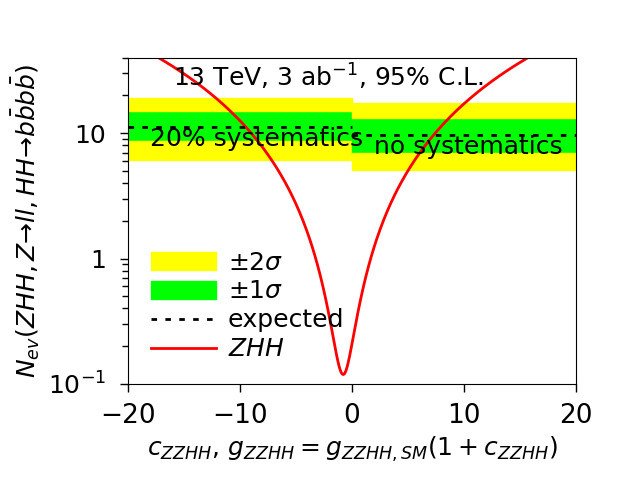}
}
\subfloat[]{
  \includegraphics[width=.49\textwidth]{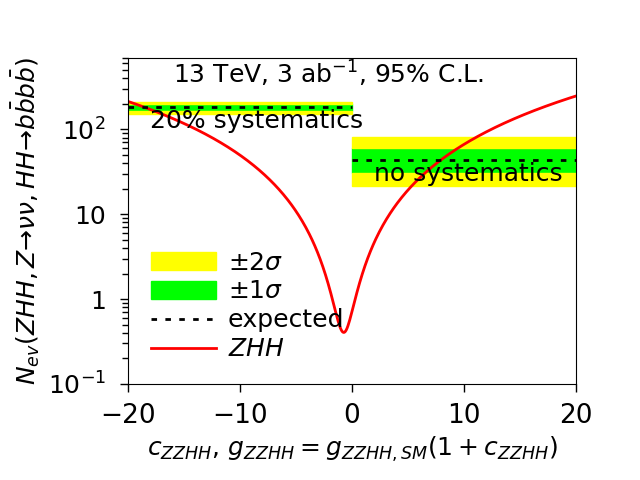}
}
\caption{Projected 95\% confidence level limits on the quartic $ZZHH$ coupling derived from the $ZHH$ analyses ($Z \to ll$ on the left, $Z \to \nu \nu$ on the right) using the full HL-LHC data set. For details on how the signal cross section and visible cross section limits are calculated see text. \label{fig:zhh_cvvhh_limits}}
\end{figure}

Our projected 95\% confidence level (C.L.) limits on $c_3$, $\lambda = \lambda_{SM}(1+c_3)$, calculated using the CLs method \cite{0954-3899-28-10-313}, for the three analyses are presented in Table~\ref{tab:c3_zhh}. We also use the same strategy to project constraints on the quartic interaction $VVHH$ and present the results in Figures~\ref{fig:zhh_cvvhh_limits}~and~\ref{fig:whh_limits_cvvhh}, with the 95\% confidence level limits on $c_{VVHH}$, $g_{VVHH} = g_{VVHH,SM}(1+c_{VVHH})$, in Table~\ref{tab:cvvhh_zhh}. Compared to the projected limits on $g_{VVHH}$ in the $HHjj$ study in \cite{Dolan:2015zja}, and in particular to the projected limits set using boosted reconstruction techniques in \cite{Bishara:2016kjn}, we see that the sensitivity of the $VHH$ production channels is quite limited. Nevertheless, the ability to separate the $ZZHH$ and $WWHH$ contributions is as noted above unique to the $VHH$ channels and these could therefore, at least in theory, provide complementary information for this measurement. The weak sensitivity here however suggests we are constraining values which violate perturbative unitarity at the scales the LHC probes: recasting the limits derived on dimension-6 operators in \cite{Corbett:2014ora} we find that values for $c_{VVHH} \sim \mathcal{O}(5)$ violate perturbative unitarity for $\sqrt{s} \sim $ TeV.

All of the results presented here assume an idealised detector with perfect performance except for the $b$-tagging (false) rates given in Section~\ref{sec:setup}. To estimate how sensitive our results are to the uncertainties introduced by a realistic detector we have also performed the same analyses using the fast simulation machinery included in \textsc{Rivet}, which allows for lepton efficiencies and jet, lepton, and missing transverse energy smearing according to reported values by ATLAS to be taken into account \cite{Aad:2012re,ATL-PHYS-PUB-2015-041}. The full results are presented in Table~\ref{tab:xstable_zhh_det} in the Appendix. The most significant effect we find is that top quark backgrounds become even more problematic for the $Z \to \nu \nu$ analysis. This reduces $S/B$ by another factor of about 10.

\begin{table}
\begin{tabular}{lccc}
\toprule
                   &    $Z \to ll$                 & $Z \to \nu \nu$  &   $W \to l \nu$    \\
No systematics     &  \hspace{0.3cm}  $-22.5 < c_3 < 19.0$ \hspace{0.3cm}      & \hspace{0.3cm} $-27.0 < c_3 < 23.1$ \hspace{0.3cm}  & \hspace{0.3cm} $-40.0 < c_{3} < 34.9$ \hspace{0.3cm}    \\
20\% systematics   &    $-23.8 < c_3 < 20.3$       & $-43.5 < c_3 < 40.0$ & --- \\
\botrule
\end{tabular}
\caption{95\% C.L. CLs limits on $c_3$, $\lambda = \lambda_{SM}(1+c_3)$, for the three analyses under hypotheses of no systematic uncertainties and 20\% systematic uncertainties. We do not include results with systematic uncertainties for the $WHH$ analysis due to its weak sensitivity. For precise definitions of how the limits are calculated see text and Figure~\ref{fig:zhh_limits}.}
 \label{tab:c3_zhh}
\end{table}

\begin{table}
\begin{tabular}{lccc}
\toprule
                   &    $Z \to ll$                 & $Z \to \nu \nu$  &   $W \to l \nu$   \\
No systematics     &  \hspace{0.3cm}  $-8.9 < c_{ZZHH} < 7.3$ \hspace{0.3cm}      & \hspace{0.3cm} $-9.4 < c_{ZZHH} < 7.9$ \hspace{0.3cm} & \hspace{0.3cm} $-11.6 < c_{WWHH} < 10.0$ \hspace{0.3cm}    \\
20\% systematics   &    $-9.5 < c_{ZZHH} < 7.9$       & $-18.5 < c_{ZZHH} < 17.0$  & --- \\
\botrule
\end{tabular}
\caption{Limits on $c_{VVHH}$, $g_{VVHH} = g_{VVHH,SM}(1+c_{VVHH})$, for the three analyses under hypotheses of no systematic uncertainties and 20\% systematic uncertainties. We do not include results with systematic uncertainties for the $WHH$ analysis due to its weak sensitivity. For precise definitions of how the limits are calculated see text and Figure~\ref{fig:zhh_cvvhh_limits}.}
 \label{tab:cvvhh_zhh}
\end{table}

\subsection{Avenues for improvement}

In light of the extremely challenging nature of this analysis that is evident from the results presented in Table~\ref{tab:xstable_zhh}, we discuss some of the improvements that we have attempted. As is evident from Figure~\ref{fig:zhh_met}, there is additional information to be used in the missing transverse energy distribution for the $Z \to \nu \nu$ analysis. We have checked explicitly that using a cut of 150 GeV leads to a small but noticeable improvement in the sensitivity. However the improvement vanishes when using a fast detector simulation due to the smearing of the $|E_T^{\textrm{miss}}|$. We have included a shape comparison for the smeared $|E_T^{\textrm{miss}}|$ distribution in the Appendix. The plain $t \bar{t}$ background could also be reduced by using an operating point for the ATLAS MV2c10 $b$-tagger with higher background rejection, however this is not a ``silver bullet'' since the $t \bar{t} b \bar{b}$ background remains sizeable and we have checked that using the most aggressive point in \cite{ATL-PHYS-PUB-2016-012} does not lead to significant change in sensitivity due to the accompanying loss of signal efficiency.

As already discussed for the $Z \to ll$ analysis, making use of lower background yields in the high-$m_{HH}$ tail of the distribution does not generate dramatic improvements. The situation for the two other analyses where $t \bar{t} + X$ backgrounds are dominant could be more promising, however any such information will be difficult to make use of effectively at the HL-LHC due to the tiny cross sections.

A multivariate approach could make use of additional information in angular distributions as discussed in Section~\ref{sec:zll}, potentially combined with observables like $m_{ll}$ and $\chi_t$ motivated by top-based backgrounds to significantly improve the $S/B$ without losing signal statistics in a way not possible in a simple cut-based analysis as presented here. However even so, the low signal rate after the basic selections we apply here will remain problematic: we have re-calculated the limit on $c_3$ from the $Z \to ll$ analysis presented in Table~\ref{tab:c3_zhh} assuming backgrounds are reduced by an order of magnitude with no cost in signal efficiency, and find $-17.5 < c_3 < 13.5$ in the absence of systematics, which illustrates that even large improvements in a multivariate analysis are unlikely to make this channel sensitive to the $c_3$ range where the cross section scaling favours it over other channels, Figure~\ref{fig:c3_scaling}.

Accessing the full phase space information through the Matrix Element method \cite{doi:10.1143/JPSJ.57.4126,Abazov:2004cs,Gao:2010qx,Artoisenet:2010cn,Alwall:2010cq,Andersen:2012kn,Artoisenet:2013vfa} or by using \textsc{MadMax} \cite{Cranmer:2006zs,Plehn:2013paa,Kling:2016lay} or similar would ultimately be necessary to make a definitive statement on the potential sensitivity of an idealised analysis, but due to the complex nature of the relevant backgrounds, this is beyond the scope of our present study. The extremely low signal rates which are evident from our simple sensitivity analysis here suggests that a differential analysis would be challenging even at a 100 TeV collider where signal cross sections are only $\sim$20 times larger.

\section{Conclusions}

Interference between the trilinear coupling-induced and and other contributions in di-Higgs production could allow subdominant channels to have competitive sensitivity to specific modifications of $\lambda$. We have investigated $VHH$ production at the HL-LHC, where this channel shows the strongest scaling with $\lambda$ for small positive modifications. Additionally, this channel could potentially be used for measuring $ZZHH$ and $WWHH$ vertices separately. Unfortunately top quark pair production backgrounds are difficult to control for the $Z \to \nu \nu$ and $W \to l \nu$ decay modes, and the signal cross section is small for the $Z \to ll$ decay mode, making the ultimate sensitivity limited. We have discussed possible avenues to improve the analysis but concluded that using $VHH$ production as a probe of modifications of the Higgs sector is likely to remain challenging even when employing more advanced techniques.

\section{Acknowledgements}

We thank Christoph Englert and Adam Falkowski for helpful comments on an earlier draft of the manuscript. We would also like to thank Kalliopi Petraki and Andy Buckley for helpful discussions. KN acknowledges support by the NWO Vidi grant "Self-interacting asymmetric dark matter". AP acknowledges support by the ERC grant ERC-STG-2015-677323.

\section{Appendix}

\begin{figure}[t]
\centering
 \includegraphics[width=.49\textwidth]{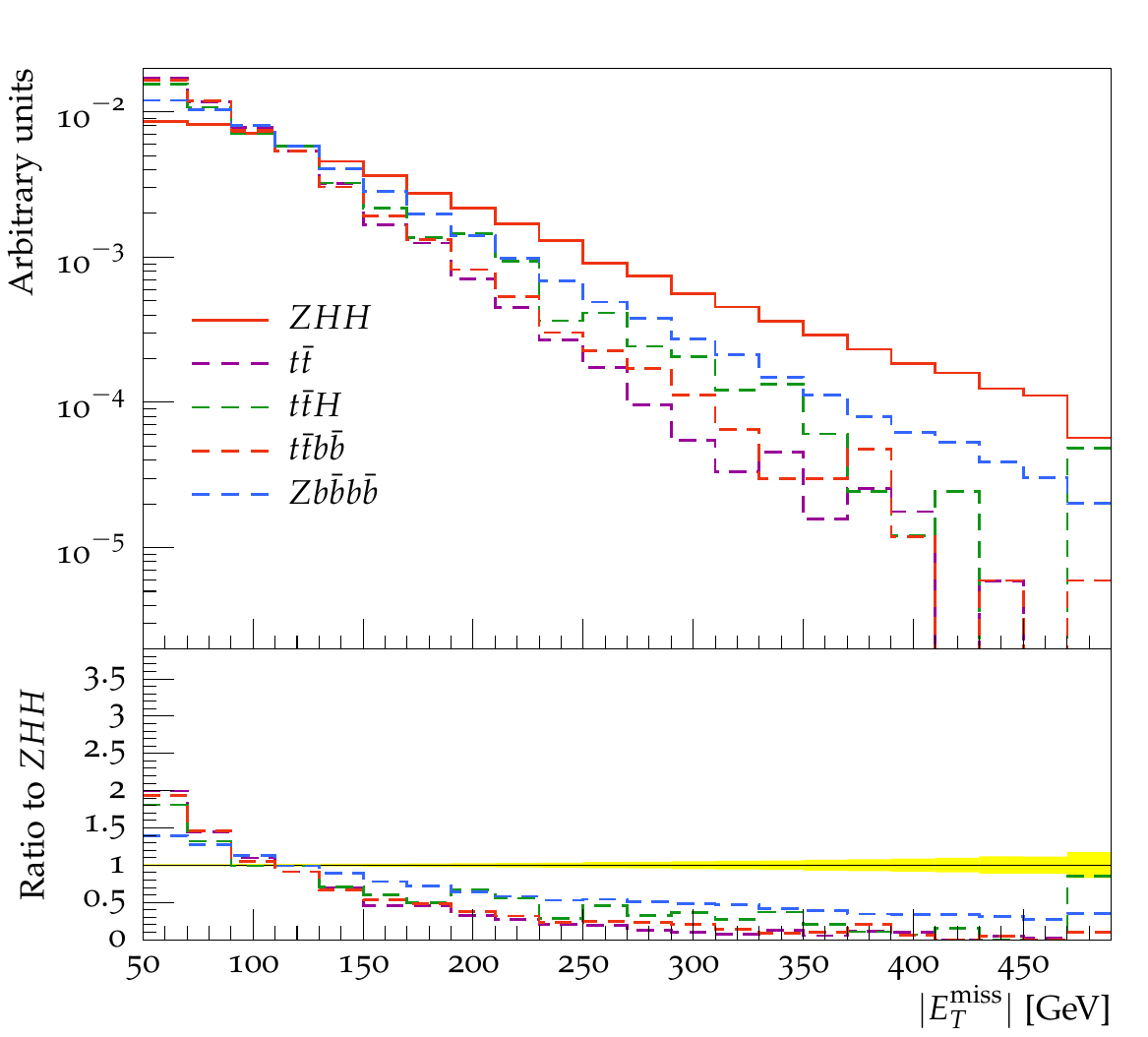}
\caption{Shape comparison of the smeared missing transverse energy distributions of the signal and leading backgrounds in the $Z \to \nu \nu$ analysis when using a fast detector simulation. \label{fig:zhh_nunu_kinematics_det}}
\end{figure}

The smeared missing tranverse energy distribution taking detector effects into account in the $Z \to \nu \nu$ analysis is presented in Figure~\ref{fig:zhh_nunu_kinematics_det}. The cross sections at different stages in the cutflow when using the fast detector simulation are provided in Table~\ref{tab:xstable_zhh_det}.

\begin{turnpage}
\begin{table*}[!t]
  \hspace{-0.5cm}  \begin{tabular}{lcccccccccc}
    \toprule
    Cut ($Z \to ll$)     		         &  $ZHH$  &  $Zb\bar{b}b\bar{b}$ &  $Zb\bar{b}c\bar{c}$ &  $Zt\bar{t}$ &  $ZZb\bar{b}$    & $t\bar{t}H$   & $t\bar{t}b\bar{b}$  & $t\bar{t}c\bar{c}$ & $t\bar{t}$ \\
    \botrule
    2 same flavour leptons	         &  \num{3.07E-06}&   \num{3.50E-03}&  \num{7.46E-02}  & 	\num{2.74E-02} & \num{2.74E-03} & \num{3.12E-03} & \num{4.47E-02}  &  \num{2.69E-01} &  \num{4.57E+01}	\\
    $|m_{ll} - m_Z| < 5$ GeV	       &  \num{1.84E-06}&   \num{1.85E-03}&  \num{5.44E-02}  &  \num{1.04E-02} & \num{1.96E-03} &	\num{1.97E-04} & \num{2.80E-03} & 	\num{1.08E-02} &  \num{1.92E+00}	\\
    $|E_T^{\textrm{miss}}| < 50$ GeV &  \num{1.77E-06}&   \num{1.79E-03}&   \num{5.41E-02} & 	\num{7.72E-03} & \num{1.94E-03} &	\num{5.20E-05} & \num{8.01E-04} & 	\num{5.21E-03} &  \num{8.02E-01}	\\
    $\ge 4$ jets with $p_T > 40$ GeV &  \num{4.65E-07}&   \num{2.49E-04}&   \num{1.29E-03} & 	\num{4.28E-03} & \num{3.02E-05} &	\num{1.48E-05} & \num{1.53E-04} & 	\num{1.51E-03} &  \num{6.57E-02}	\\
    4 leading jets $b$-tagged	       &  \num{4.55E-08}&   \num{1.95E-05}&   \num{2.41E-06} & 	\num{3.23E-06} & \num{1.44E-06} & \num{9.17E-07} & \num{4.34E-06} & 	\num{2.14E-07} &  \num{5.23E-06}	\\
    $\chi_{HH} < 1.6$	               &  \num{2.60E-08}&   \num{1.41E-06}&   \num{2.35E-07} & 	\num{7.11E-07} & \num{1.42E-07} &	\num{3.38E-08} & \num{9.95E-07} & 	\num{6.41E-08} &  \num{1.16E-07}	\\
    $\Delta \eta (Z, H_1) < 2$	     &  \num{2.55E-08}&   \num{1.04E-06}&   \num{1.09E-07} & 	\num{5.80E-07} & \num{1.31E-07} &	\num{3.38E-08} & \num{9.95E-06} & 	\num{6.41E-07} &  \num{1.16E-07}	\\
    \botrule
    Events in 3 ab$^{-1}$            & \num{0.07660}     & \num{3.119}  & \num{0.3262}   & \num{1.739}      & \num{0.3929}    &  \num{0.1014} & \num{2.985}      & \num{0.1923}  &  \num{0.3467}
    \\
    \\
    \\
    \toprule
    Cut ($Z \to \nu \nu$)     		   &  $ZHH$   &  $Zb\bar{b}b\bar{b}$ &  $Zb\bar{b}c\bar{c}$ &  $Zt\bar{t}$&  $ZZb\bar{b}$ & $t\bar{t}H$  & $t\bar{t}b\bar{b}$   & $t\bar{t}c\bar{c}$ & $t\bar{t}$  \\
    \botrule
    No identified leptons         	 &  \num{1.45E-05}& \num{1.30E-02} &  \num{3.58E-01}    &\num{4.14E-02} &\num{1.28E-02} &\num{9.25E-02}& 	\num{1.04E+01} & 	\num{1.70E+00} & 	\num{3.79E+02} 	\\
    $|E_T^{\textrm{miss}}| > 100$ GeV&  \num{6.86E-06}& \num{4.20E-03}&   \num{5.76E-02}    &\num{4.01E-03} &\num{1.97E-03} &\num{8.32E-03}& 	\num{7.24E-01} & 	\num{1.30E-01} & 	\num{2.44E+01} 	\\
    $\ge 4$ jets with $p_T > 40$ GeV &  \num{2.27E-06}& \num{9.08E-04} &  \num{4.30E-03}    &\num{2.82E-03} &\num{1.27E-04} &\num{6.35E-03}& 	\num{3.02E-01} & 	\num{7.91E-02} & 	\num{8.48E+00} 	\\
    4 leading jets $b$-tagged	       &  \num{2.21E-07}& \num{6.95E-05} &  \num{7.53E-06}    &\num{4.95E-05} &\num{6.68E-06} &\num{9.51E-05}& 	\num{1.78E-03} & 	\num{1.08E-04} & 	\num{2.66E-03} 	\\
    $\chi_{HH} < 1.6$	               &  \num{1.23E-07}& \num{3.73E-06} &  \num{3.14E-07}    &\num{3.01E-06} &\num{7.47E-07} &\num{9.32E-06}& 	\num{2.12E-04} & 	\num{8.51E-06} & 	\num{6.43E-04} 	\\
    \botrule
    Events in 3 ab$^{-1}$            & \num{0.3677}   & \num{11.19}     & \num{0.9416}       & \num{9.043}   & \num{2.241}   & \num{27.97}  &  \num{637.4}    &  \num{25.52}    &  \num{1929}
    \\
    \\
    \\
    \toprule
Cut ($W \to l \nu$)                          &  $WHH$  &  $t\bar{t}$ & $t\bar{t}b\bar{b}$ & $t\bar{t} H$ & $t\bar{t}Z$ &  $W b\bar{b}b\bar{b} $ & $W t\bar{t}$  \\
\botrule
$\ge 4$ jets with $p_T > 40$ GeV & \num{1.79932168e-05} & \num{151.045385493} & \num{0.58092350843} & \num{0.0668271669217} & \num{0.011589540631} & \num{0.00234876695863} & \num{0.030270941576} \\
Exactly one lepton & \num{1.1941572e-05} & \num{104.267149136} & \num{0.416557467561} & \num{0.0485375212096} & \num{0.00846042045329} & \num{0.00151302257568} & \num{0.0202407045222} \\
$|E_T^{\textrm{miss}}| < 40$~GeV & \num{3.60889688e-06} & \num{14.7706179619} & \num{0.1583682636} & \num{0.0224883060206} & \num{0.00334320995751} & \num{0.000264635672091} & \num{0.00716817343465} \\
$m_T < 80.42$~GeV and $H_T > 400$~GeV   & \num{1.96944288e-06} & \num{10.054931356} & \num{0.130579208351} & \num{0.0195272478423} & \num{0.0027895522908} & \num{0.000100247297549} & \num{0.00595873402108} \\
4 leading jets $b$-tagged & \num{1.22751592e-07} & \num{0.000328167025909} & \num{0.00212354063269} & \num{0.000315998940707} & \num{5.10254752407e-05} & \num{6.76533143735e-06} & \num{1.93462190622e-06} \\
$\chi_{HH} < 1.6$ & \num{7.1877664e-08} & \num{4.30989780631e-05} & \num{9.98841359781e-05} & \num{3.58297572948e-05} & \num{2.72668969176e-06} & \num{2.52316573506e-07} & \num{2.84888691506e-07} \\
\botrule
Events in 3~ab$^{-1}$ & \num{0.215632992} & \num{129.296934189} & \num{299.652407934} & \num{107.489271885} & \num{8.18006907529} & \num{0.756949720518} & \num{0.854666074517}
     \end{tabular}
     \caption{Cross sections in picobarns for $ZHH$ ($Z \to ll$ and $Z \to \nu \nu$) and $WHH$ ($W \to l \nu$), and backgrounds after the selections described in the text are applied when using the fast detector simulation included in \textsc{Rivet}. Generation level cuts on the invariant masses of lepton pairs, missing transverse energy, and $p_T$ of jets are employed for some of the backgrounds. Top quark branchings depend on the analysis to allow for the possibility of leptons escaping detection, and to ensure there are two leptons for the $Z \to ll$ analysis. The $ZHH$, $WHH$, $Zt\bar{t}$, $t \bar{t}H$, $Wt\bar{t}$ and $t \bar{t}$ samples are generated at NLO QCD. The $t \bar{t}$ sample is further reweighted to the NNLO+NNLL QCD cross section \cite{Czakon:2011xx}. The other samples are generated at leading order and the $t\bar{t}b\bar{b}$ sample is reweighted to the NLO QCD cross section \cite{Bredenstein:2009aj,Bevilacqua:2009zn}. }
 \label{tab:xstable_zhh_det}
\end{table*}

\end{turnpage}

\bibliography{bibfile}

\end{document}